\documentclass[final,5p,times,twocolumn,english,fleqn]{elsarticle}
\pdfoutput=1
\usepackage{amssymb}
\usepackage{babel}
\usepackage{graphicx}
\usepackage{amsmath,amssymb,amsfonts,threeparttable}
\usepackage{textcomp}
\usepackage{gensymb}
\usepackage{booktabs}
\usepackage{hyperref}
\hypersetup{
    breaklinks=true,
    unicode=false,          
    pdftoolbar=true,        
    pdfmenubar=true,        
    pdffitwindow=false,     
    pdfstartview={FitH},    
    pdfauthor={Venkat},     
    pdfsubject={ABL},   
    pdfcreator={Venkat},   
    pdfproducer={Venkat}, 
    pdfnewwindow=true,      
    colorlinks=true,       	
    linkcolor=black,          
    citecolor=black,    	    
    filecolor=black,   		    
    urlcolor=blue            
}
\journal{Journal of Magnetism and Magnetic Materials}
\usepackage[labelfont=bf]{caption}
\makeatletter
\def\ps@pprintTitle{%
 \let\@oddhead\@empty
 \let\@evenhead\@empty
 \def\@oddfoot{}%
 \let\@evenfoot\@oddfoot}
\makeatother

\makeatletter
\def\blfootnote{\gdef\@thefnmark{}\@footnotetext}
\makeatother

\begin{document}

\title{Absorbing boundary layers for spin wave micromagnetics}

\author[label1]{G. Venkat\corref{mycorrespondingauthor}}
\cortext[mycorrespondingauthor]{Corresponding author}
\ead{guruvenkat7@gmail.com}
\author[label2,label3]{H. Fangohr}
\author[label1]{A. Prabhakar}

\address[label1]{Dept. of Electrical Engineering, Indian Institute of Technology Madras, India 600036}
\address[label2]{Faculty of Engineering and the Environment, University of Southampton, UK.}
\address[label3]{European XFEL GmbH, Holzkoppel 4, 22869 Schenefeld, Germany}

\begin{abstract}
Micromagnetic simulations are used to investigate the effects
of different absorbing boundary layers (ABLs) on spin waves (SWs)
reflected from the edges of a magnetic nano-structure. We define the
conditions that a suitable ABL must fulfill and compare the performance of
abrupt, linear, polynomial and tan hyperbolic damping profiles in the ABL. We
first consider normal incidence in a permalloy stripe and propose a 
transmission line model to quantify reflections and calculate the loss 
introduced into the stripe due to the ABL. We find that a parabolic damping profile absorbs
the SW energy efficiently and has a low reflection coefficient, thus performing much
better than the commonly used abrupt damping profile. We then investigated SWs that are obliquely  incident
at $26.6^{\circ}$, $45^{\circ}$ and $63.4^{\circ}$ on the edge of a yttrium-iron-garnet film. The parabolic damping profile
again performs efficiently by showing a high SW energy transfer to the ABL and a low reflected SW amplitude.
\end{abstract}


\maketitle

\blfootnote{ABL - Absorbing Boundary Layer; SW - Spin Wave; LL - Landau-Lifshitz; PML - Perfectly Matched Layer; FDTD - Finite Difference
Time Domain; GPU - Graphics Processing Unit; FD - Finite Difference; YIG - Yttrium Iron Garnet; Transmission line - Tx line}

\section{Introduction}

Easier access to computational resources over the
last decade has led to the development of many micromagnetic packages that
solve the Landau-Lifshitz (LL) equation for magnetic nano-structures. 
These packages are being used to study spin wave mode profiles and spectra in a quest
to build devices with novel functionalities \cite{Bance/Schrefl/Hrkac/2008, Peng201757, Magnonics_Review}.
One approach to these studies is to perturb the ground state with
a broadband excitation, and then extract the spin wave (SW) dispersion 
characteristics \cite{SKKim/2010, Li201649, silvani2017spin, standard_micromagnetic_problem_Sw_dispersion}.
However, simulation boundaries are known to affect 
the dissipative dynamics of the magnonic spectra in such 
studies \cite{Role_bound_2014, Role_of_boundary_2013}, and we artificially
increase the damping $\alpha$ at the boundaries, to absorb the SW reflections. The increase in $\alpha$ can be smooth, e.g. using 
a hyperbolic tangent function \cite{berkov:08Q701}, or abrupt \cite{Consolo/abrupt_damping/2007}. 
The latter approach was used to attenuate SW reflections, and to calculate 
the dispersion and scattering parameters in magnonic devices \cite{Dvornik11a, SW-abrupt-damping-1}.
More recently, an exponential increase in damping was used to curb 
reflections in the study of skyrmions and the Dzyaloshinskii-Moriya
interaction in magnetic nanostripes \cite{Zhang_2015, Xia_2016}.

\par In this article, we define the return loss using transmission line models,
to study the impact of using artificial regions of high $\alpha$,
or absorbing boundary layers (ABLs), at the edges of the device. 
We propose a parabolic increase in $\alpha$ and show that it causes less spurious SW reflections than an abrupt
increase in $\alpha$. We compare the parabolic profile against the abrupt,
linear and the tan hyperbolic profile, for different angles of incidence.
The parabolic profile also aligns the micromagnetic community more closely with the accepted
polynomial form of perfectly matched layers (PMLs) in finite difference time domain (FDTD)
simulations of Maxwell's equations \cite{PML_criteria}.

\par To our knowledge, this is the first exhaustive study of ABLs using the graphics processing unit (GPU) accelerated finite difference (FD) micromagnetic  
package MuMax3 \cite{MuMax3}. We also provide the codes for post processing the
simulation data and raw data for the figures in a code repository for easy reproduction \cite{Git_codes}.

\section{Normal incidence of spin waves}\label{sec:normal-incidence}

\begin{figure}[th]
\centering{}{\includegraphics[width=0.9\columnwidth]{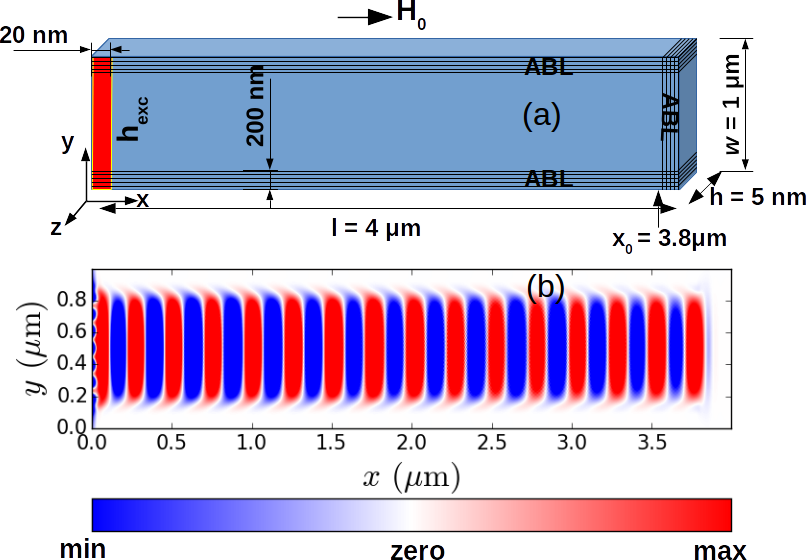}
\protect\caption{ \label{fig:Fig-1} (a) A magnonic waveguide with absorbing
boundary layers along all edges. A SW excitation pulse ${\bf h_{\text{exc}}}(t)$,
is applied along $\hat{\boldsymbol{y}}$, at the left edge. The origin is at the 
bottom left corner. (b) A snapshot of $m_y$ at $t=500\,\text{ns}$. The colorbar is
in linear scale.}
}
\end{figure}

The time evolution of the magnetization is described by the LL equation \cite{Landau35, Franchin09}
\begin{equation}
  \begin{aligned}
    \frac{\partial{\bf m}}{\partial t} = \gamma^{\prime}\left[\left({\bf m\times{\bf H}}\right)+\alpha\left({\bf m}\times({\bf m}\times{\bf H})\right)\right],\label{eq:1}
  \end{aligned}
\end{equation}
where $\mathbf{m}=\mathbf{M}/M_{\text{S}}$ is the normalized magnetization,
and $\mathbf{M}$ and $\mathbf{H}$ are the total magnetization
and effective field at time $t$, respectively. $\gamma^{\prime}=\gamma\mu_{0}/(1+\alpha^{2})$, 
with $\gamma<0$ being the electron gyromagnetic ratio, $\alpha$ the phenomenological
damping coefficient and $\mu_{0}$ the permeability of free space. We consider
a stripe of permalloy ($\text{Ni}_{80}\text{Fe}_{20}$) having dimensions
$4000\times1000\times5\,{\text{nm}}^3$, as shown in  \autoref{fig:Fig-1} (a).
The structure was proposed as a micromagnetic sample problem for studying SW dynamics and dispersion \cite{standard_micromagnetic_problem_Sw_dispersion}.
We choose a simple geometry with known solutions for the mode profiles.

The material parameters used for permalloy were the saturation magnetization $M_{\text{\text{{s}}}} = \text{{800}}\,$kA/m and
exchange constant $A=\text{{13\ensuremath{\times}1\ensuremath{0^{-12}}}}$\,J/m \cite{standard_micromagnetic_problem_Sw_dispersion}. No crystalline anisotropy was considered.
The cell size was taken as $4\times4\times5\,{\text{nm}}^3$,
such that the cell dimensions are less than the exchange length for
permalloy, $l_{\text{ex}}\simeq5.7\,\text{nm}$.

\begin{figure}[tbh]
\begin{centering}
\includegraphics[width=0.9\columnwidth]{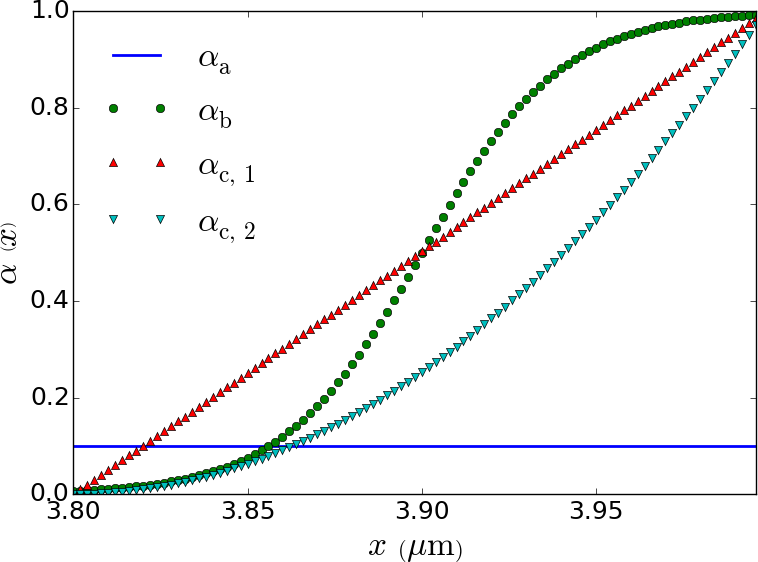}{\protect\caption{{ \label{fig:Fig-2}
Spatial variation of different damping profiles that were studied. $\alpha_{c,2}$ is the 
parabolic damping profile.}
}}
\par\end{centering}

\end{figure}

A harmonic field excites SWs at the left edge of the stripe so that they propagate along $\hat{\boldsymbol{x}}$.
ABLs along the top and bottom stripe edges confine the SWs to the centre of the stripe, as shown in  \autoref{fig:Fig-1} (b). 
Now, consider different spatial profiles for damping, defined at the right end as:
\begin{itemize}
\item {constant and abrupt}\\
{$\alpha_{\text{a}}\left(x\right)=\begin{cases}
0 & x<3.8\,\text{\ensuremath{\mu}m}\\
0.1 & x\geq3.8\,\text{\ensuremath{\mu}m}
\end{cases}$}\\

\item tan hyperbolic, with $\Delta\alpha=0.5$, $x^{\prime}=3.9\,\text{\ensuremath{\mu}m}$
and $\sigma_{x}=40\,\text{nm}$, modified from \cite{berkov:08Q701}\\
\\
{$\alpha_{\text{b}}\left(x\right)=\begin{cases}
0 & x<3.8\,\text{\ensuremath{\mu}m}\\
\Delta\alpha(1+\tanh\frac{x-x^{\prime}}{\sigma_{x}}) & x\geq3.8\,\text{\ensuremath{\mu}m}
\end{cases}$}\\

\item {polynomial, with $x_{0}=3.8\,\text{\ensuremath{\mu}m}$}\\
{}\\
{$\alpha_{\text{c},n}\left(x\right)=\begin{cases}
0 & x<3.8\,\text{\ensuremath{\mu}m}\\
a(x-x_{0})^{n} & x\geq3.8\,\text{\ensuremath{\mu}m}\quad n=1,2
\end{cases}$}\\

\end{itemize}
In each case the constants were chosen to obtain $\alpha = 1.0$ at $x=4\,\mu \text{m}$, as shown in  \autoref{fig:Fig-2}.
$\alpha_{\text{c},1}$ and $\alpha_{\text{c},2}$ are linear and parabolic profiles respectively. We compare different profiles
over a constant ABL length of $200\,\text{nm}$. In the following sections, we also show that $200\,\text{nm}$ is
sufficient for the energy density to decay by over $15\,\text{dB}$, for all the damping profiles.

{}
\begin{figure}[tbh]
\begin{centering}
\includegraphics[width=0.9\columnwidth]{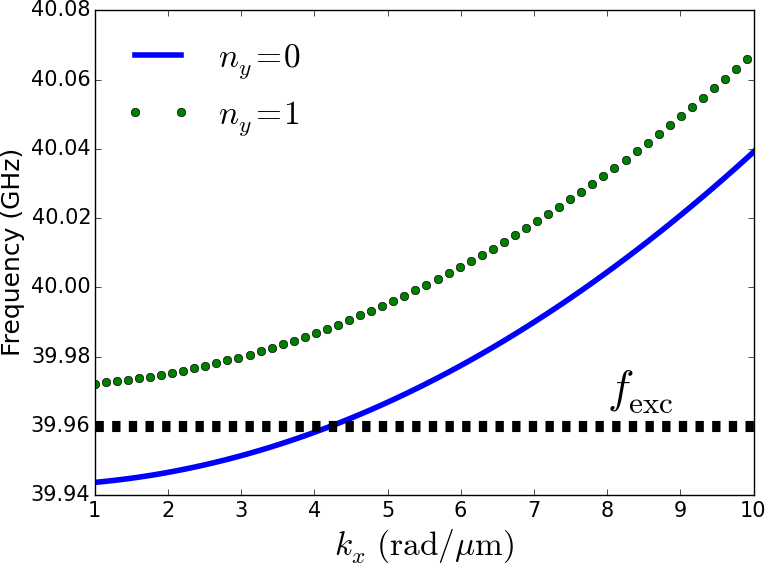}{{\protect\caption{\label{fig:Fig-3}{SW dispersion in the magnonic stripe, showing how
$f_{\text{exc}}=39.96\,\text{GHz}$ excites only the fundamental mode.}}
}}
\par\end{centering}

\end{figure}

\subsection{Simulation procedure}

We apply a high bias magnetic field $\mathbf{H}_{0}=804\,\text{kA/m}\,\hat{\boldsymbol{x}}$, with an artificially high damping $\left(\alpha=0.5\right)$,
and allow $\mathbf{m}$ to relax to its ground state. 
Since the magnetization in the stripe is saturated, we do not have any domain
walls or vortices in the stripe. Starting with the ground state, an excitation 
magnetic field
\begin{eqnarray}
  \begin{aligned}
  \mathbf{h}_{\text{exc}}\left(x,\,y,\,t\right) & = & h_{0}\sin\left(2\pi f_{\text{exc}}t\right)\cos\left(\frac{\pi}{2w}y-\frac{\pi}{4}\right)\hat{\boldsymbol{y}},\label{eq:2}
  \end{aligned}
\end{eqnarray}
is applied at $x<20\,\text{nm}$ (in the region marked in red in \autoref{fig:Fig-1} (a)) with $h_{0}=0.01H_{0}$, $f_{\text{exc}}=39.96\,\text{GHz}$
and the width of the stripe $w=1\,\text{\ensuremath{\mu}m}$. A low
value of $h_{0}$ ensures that we excite small amplitude SWs. The
spatial form of $\cos\left(\frac{\pi}{2w}y-\frac{\pi}{4}\right)$
was chosen so that we preferentially excite the lowest order width mode.

The dispersion relation for the lowest SW mode in a backward volume geometry 
$(\mathbf{k} \parallel \mathbf{H}_0)$ was 
derived by Kalinikos \cite{Kalinikos/1980}. If we include exchange interactions, we get
\begin{equation}
	\begin{aligned}
		& \omega  =  \sqrt{\omega_{\text{ex}}\left(\omega_{\text{ex}}+\omega_{\text{M}}\frac{1-e^{-kh}}{kh}\right)}, \\
		& \omega_{\text{ex}}  =  \omega_{0} + \omega_{\text{M}}\lambda_{\text{ex}}k^2,\label{eq:3}
	\end{aligned}	
\end{equation}
where $h$ is the stripe thickness, $\omega_{0}=\gamma \mu_{0} H_{0}$ is the uniform mode precession frequency and $\omega_{\text{M}}=\gamma \mu_{0} M_{\text{S}}$.
$\lambda_{\text{ex}} = \frac{2A}{\mu_{0}M_{\text{s}}^{2}}$ where $A$ is the exchange constant. $k^{2}=k_{x}^{2}+k_{y}^{2}$ where $k_x$
is the propagation constant and $k_{y}=\left(n_{y}+1\right)\frac{\pi}{w}$ is the quantized
wave vector component along the width. We choose $n_{y}=0$, and pick $f_\text{exc}=39.96\,\text{GHz}$, to excite 
only the fundamental mode, as shown in  \autoref{fig:Fig-3}.
$\mathbf{m}\left(x,\,y,\,z,\,t\right)$ is saved at all the nodes
of the FD grid. The SWs take approximately $25\,\text{ns}$ to reach
the right end of the stripe. We allow the simulation to run
till $500\,\text{ns}$ so that the SWs travel ten round trips in the stripe.


\subsection{Transmission line model for ABLs}
The purpose of an ABL is three fold:
\begin{enumerate}
\item The SWs should decay sufficiently by the end of the ABL to have no reflections from the stripe edge.
\item The ABL causes minimum reflections back into the device.
\item Minimum energy is reflected into higher order modes.
\end{enumerate}
Consequently, we evaluate the different ABLs, using as a metric the energy density in the ABL and reflections from the ABL.

\par
\begin{figure}[tbh]
\centering{}\includegraphics[width=0.9\columnwidth]{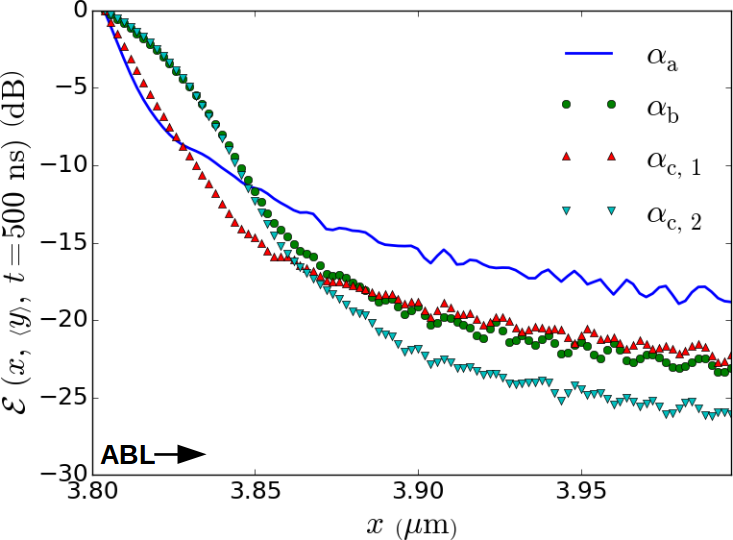}\protect\caption{ \label{fig:Fig-4} The decay of the normalized SW energy density $\mathcal{E}$ in the ABL.
The energy decays by more than $15\,\text{dB}$ within $200\,\text{nm}$ for all profiles.}
\end{figure}

 The energy density of the SWs propagating along the stripe is \cite{MuMax3}
\begin{eqnarray}
  \begin{aligned}
    \mathcal{E}\left(x,\,y,\,t \right) & = & -\frac{1}{2}\mathbf{M}\left(x,\,y,\,t \right).\mathbf{B}\left(x,\,y,\,t \right),\label{eq:5}
  \end{aligned}  
\end{eqnarray}
where $\mathbf{B}$ is the instantaneous magnetic flux density.  \autoref{fig:Fig-4} shows the variation of the normalized energy density in the ABL at
$t=500\,\text{ns}$. $\mathcal{E}$ decays by over $15\,\text{dB}$ within $200\,\text{nm}$ for all the profiles.
We observe no significant reflections
from the structure edge, and hence we fix our ABL length at $200\,\text{nm}$ for all the profiles.

\par
\begin{figure}[tbh]
\centering{}\includegraphics[width=0.9\columnwidth]{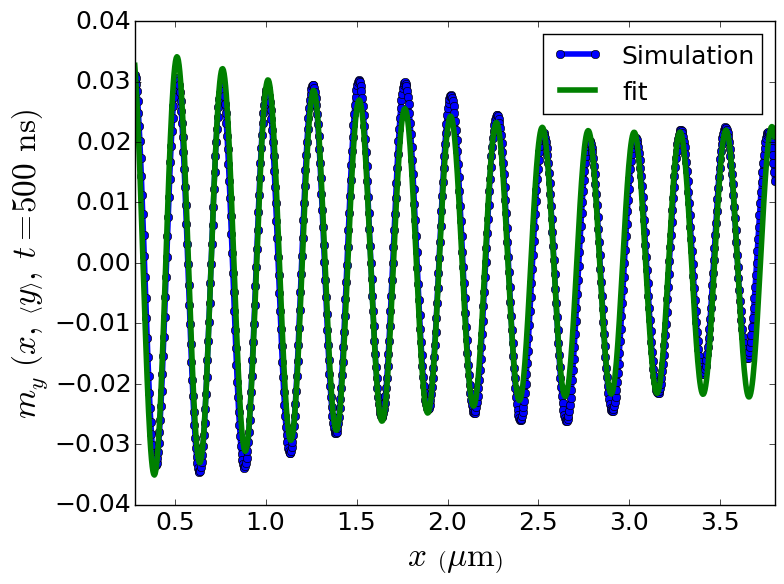}\protect\caption{ \label{fig:Fig-5} A fit of Eq. (\autoref{eq:6}) with the magnetization
in the stripe at $t_{0}=500\,\text{ns}$, when the $\alpha_{\text{c},2}$ profile is used
in the ABL. The fit is used to estimate the return loss appearing in the
line due to the introduction of the ABL.}
\end{figure}

 We now investigate reflections that originate from the start of the
ABL at $x=3.8\,\text{\ensuremath{\mu}m}$. When we make a transition from $\alpha = 0 $ to $\alpha \neq 0$, we observe SW reflections
in a manner analogous to having an impedance mismatch along a transmission (Tx) line.
We model the wave propagation in the stripe as standing waves formed on a lossy Tx line. 
In its simplest form, the magnetization on this line takes the form (c.f. Appendix A)
\begin{equation}
  \begin{aligned}
    m\left(x\right)=m^{+}\left[e^{-\zeta x}\cos\beta x+\left|\Gamma\right|e^{+\zeta x}\cos\left(\beta x+\phi\right)\right],\label{eq:6}
  \end{aligned}  
\end{equation}
where $m^{+}$ is the peak amplitude of the incident wave, $\zeta$
is the loss per unit length, $\beta$ is the propagation constant of the standing
wave and $\Gamma=\left|\Gamma \right|e^{j\phi}$ is the complex reflection coefficient at the load
end. We fit the standing wave $m_y\left(x,\,\left\langle y\right\rangle ,\,t\right)$, in the stripe, to
Eq. (\autoref{eq:6}) for each of the different damping profiles. These fits are done for $t=475$ to $t=500\,\text{ns}$, to obtain the mean and standard deviation for
$\zeta$, $|\Gamma|$ and $\phi$. One such fit is shown in  \autoref{fig:Fig-5} for $\alpha_{\text{c},2}$ at $t_{0}=500\,\text{ns}$.



The return loss in a Tx line is a measure of the power reflected
by a mismatched load and is given as \cite{pozar1997microwave}
\allowdisplaybreaks
\begin{eqnarray}
  \begin{aligned}
    \text{RL} & = & -20\log_{10}\left|\Gamma\right|\,\text{dB}.\label{eq:7}
  \end{aligned}
\end{eqnarray}
The time averaged $\zeta$ 
and $\text{RL}$ values (along with the precision) are given for the different profiles in Table. \ref{tab:RL}.
A higher value of RL indicates a lower reflection coefficient and thus a more matched load, and
the parabolic profile shows a 1.5\,\text{dB}
higher $\text{RL}$ than the commonly used abrupt profile. The value of 
$\text{RL}$ for the parabolic profile ($\alpha_{\text{c},2}$)  is comparable to that of the 
tan hyperbolic profile ($\alpha_{\text{b}}$) and therefore both appear to be efficient for use in an ABL.

\begin{table}[tbh]
\centering
\begin{threeparttable}
\protect\caption{\label{tab:RL}$\zeta$ and Return loss for the different ABL profiles.}
\begin{tabular}{l l l l}
\toprule
S. No. & Profile & $\zeta\,\left(\mu \text{m}^{-1} \right)$ & $\text{RL}\,\left(\text{dB}\right)$ \tabularnewline
\midrule
1 & $\alpha_{\text{a}}$ & $0.08\pm 0.01$ & $5.21\pm0.01$\tabularnewline

2 & $\alpha_{\text{b}}$ & $0.1\pm0.02$ & $6.99\pm0.02$\tabularnewline

3 & $\alpha_{\text{c},1}$ & $0.08\pm0.01$ & $5.41\pm0.02$\tabularnewline

4 & $\alpha_{\text{c},2}$ & $0.1\pm0.02$ & $6.72\pm0.01$\tabularnewline
\bottomrule 
\end{tabular}
\end{threeparttable}
\end{table}

\section{Oblique incidence of spin waves}

\begin{figure}[tbh]
\centering{}\includegraphics[width=0.9\columnwidth]{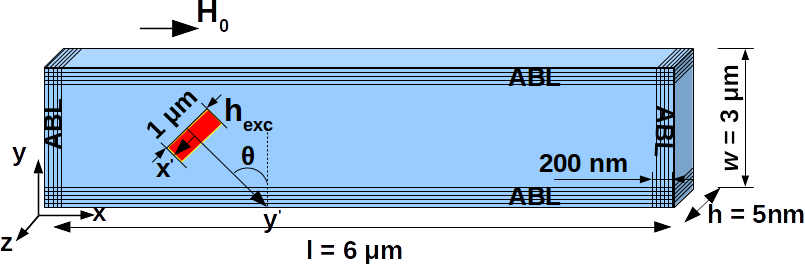}\protect\caption{ \label{fig:Fig-6}A thin film of YIG with ABLs all around
the periphery and with an excitation region at an angle. The origin is at the bottom left corner.}
\end{figure}

In FDTD simulations, the performances of PMLs are typically functions
of the angles at which the electromagnetic waves are incident on them. Having shown
the performance of ABLs for perpendicular incidence in \autoref{sec:normal-incidence},
we now investigate their effect when we have oblique incidence. Consider
the geometry recently used to simulate the Goos-Hanchen effect for
SWs \cite{Goos-Hanchen-SWs}, in a yttrium iron garnet (YIG) film, which is shown in 
\autoref{fig:Fig-6}. The dimensions of the film are $6000\times3000\times5\,{\text{nm}}^3$.
The material parameters used for YIG were $M_{\text{\text{{s}}}} = \text{{194}}
\,$kA/m and $A=\text{{4\ensuremath{\times}1\ensuremath{0^{-12}}}}$\,J/m \cite{Goos-Hanchen-SWs}. Again no crystalline anisotropy was considered.
{}

\begin{figure}[tbh]
\centering{}\includegraphics[width=1.0\columnwidth]{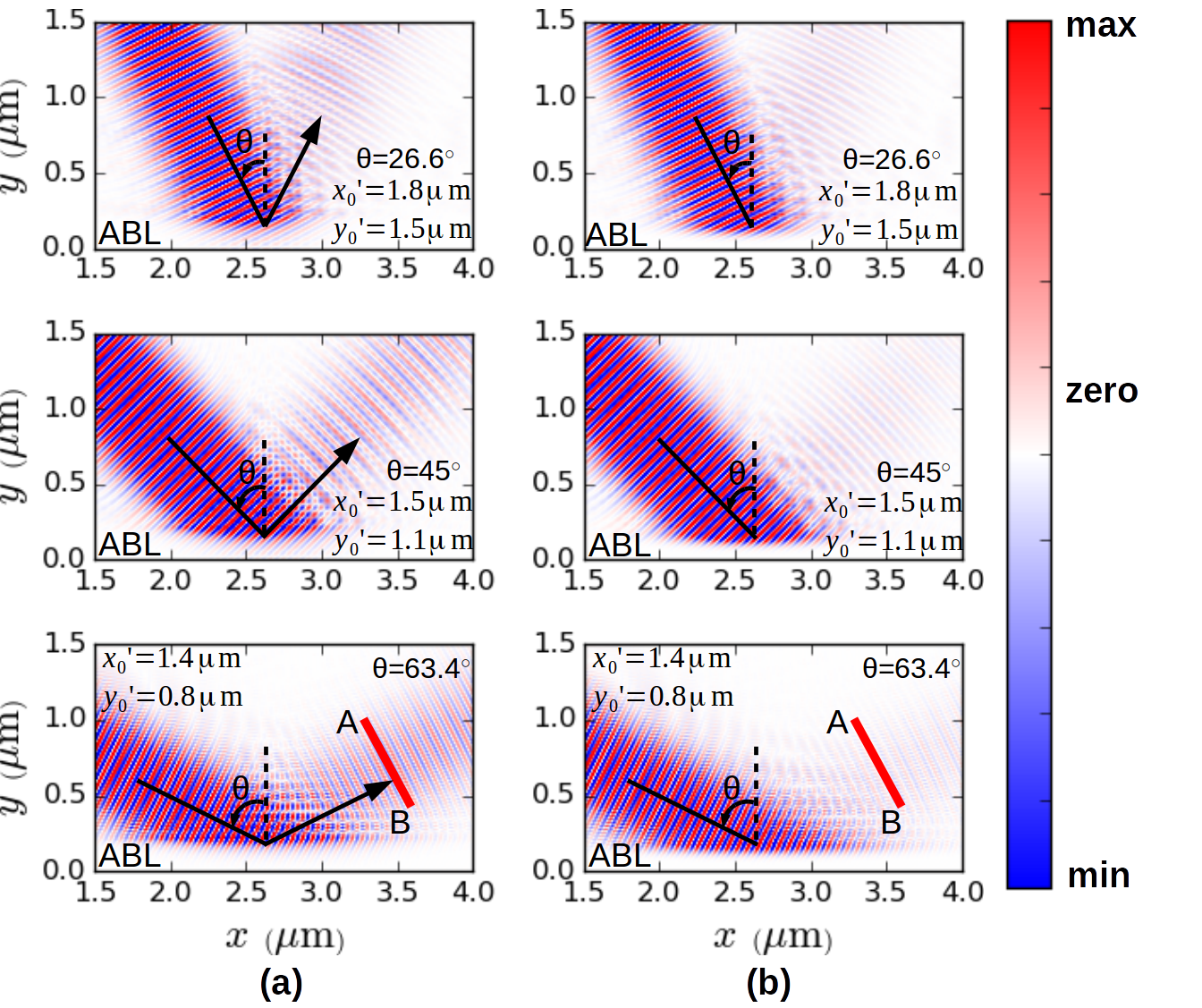}\protect\caption{
 \label{fig:Fig-7}SWs in the YIG film for different incident angles.
Columns (a) and (b) have the  $\alpha_{\text{a}}$ and $\alpha_{\text{c},2}$
profiles in the ABL respectively. The colorbar is in linear scale. $\left(x_{0}^{\prime},\,y_{0}^{\prime}\right)$ is the centre of the excitation region and 
was appropriately chosen for each excitation angle. The magnetization is scanned along the wavefront (red line) to obtain the plot in  \autoref{fig:Fig-9}. $\alpha_{\text{c},2}$
causes minimal reflections for all three angles of incidence. }
\end{figure}

The $\alpha_{\text{c},2}$
profile was applied along the left, top and right edges allowing us to focus on reflections off the bottom
edge. We apply a magnetic field $\mathbf{H}_{0}=558\,\text{kA/m}\,\hat{\boldsymbol{x}}$
and allow $\mathbf{m}$ to relax to its ground state. We then choose an area at an angle $\theta$, as shown
in  \autoref{fig:Fig-6}, and apply \cite{Goos-Hanchen-SWs}
\begin{eqnarray}
  \begin{aligned}
    \mathbf{h}_{\text{exc}}\left(x^{\prime},\,y^{\prime},\,t\right)=h_{0}e^{-2\left(\frac{x^{\prime}-x_{0}^{\prime}}{l_{\text{exc}}\sigma_{\text{exc}}}
    \right)^{2}}\sin\left(2\pi ft\right)\hat{\boldsymbol{y}},\label{eq:8}
  \end{aligned}
\end{eqnarray}
with $h_{0}=0.01H_{0}$. 
$\hat{\boldsymbol{y}}$ is the desired direction of SW propagation, at an angle $\theta$, and $\hat{\boldsymbol{x}}^{\prime}$ 
is the direction of spin wavefronts. $\left(x_{0}^{\prime},\,y_{0}^{\prime}\right)$ marks the centre
of the excitation region, and was chosen appropriately for the different angles of incidence considered below, and shown
in Fig. \ref{fig:Fig-7}. The choice of $\left(x_{0}^{\prime},\,y_{0}^{\prime}\right)$ ensured that point of 
incidence was the same for each simulation.

$l_{\text{exc}}=1\,\text{\ensuremath{\mu}m}$ and $w_{\text{exc}}=5\,\text{nm}$
are the length and width of the excitation area, and we apply $\mathbf{h}_{\text{exc}}$ to all mesh nodes that fall within this region. 
$\sigma_{\text{exc}}=0.4$ decides the spread of the Gaussian envelope.
We tested the ABL for sinusoidally pumped spin waves with $f=35\,\text{GHz}$
\cite{Goos-Hanchen-SWs}.

We observed that proper SW collimation was obtained when the SW propagation angle 
$\left(\theta \right)$ was related to the cell edge lengths, $\Delta x$ and $\Delta y$, by
$\tan \theta = \frac{\Delta y}{\Delta x}$. Consequently, we considered three cases where we took $\Delta x = 5\,\text{nm}$ and
$\Delta y = 2.5,\,5\,\text{and}\,10\,\text{nm}$. Each of these edge lengths is smaller than the exchange length
of YIG ($l_{\text{ex}}\approx13\,\text{nm}$). For these three cases, $\tan \theta = 0.5,\,1\,\text{and}\,2$ which lead to
$\theta=26.6^{\circ},\,45^{\circ}\,\text{and}\,\,63.4^{\circ}$ respectively.

The snapshots for the $\alpha_{\text{a}}$ and $\alpha_{\text{c},2}$ profiles,
for the different $\theta$, are shown in  \autoref{fig:Fig-7}. We see
significant reflections when $\alpha_{\text{a}}$ is used whereas $\alpha_{\text{c},2}$
hardly shows any reflections for the three angles of incidence. The larger reflections from $\alpha_{\text{a}}$ leads to regions of constructive and destructive
interference close to the point of incidence.
Such artifacts are avoided with $\alpha_{\text{c},2}$.
\par
\begin{figure}[tbh]
\centering{}\includegraphics[width=0.9\columnwidth]{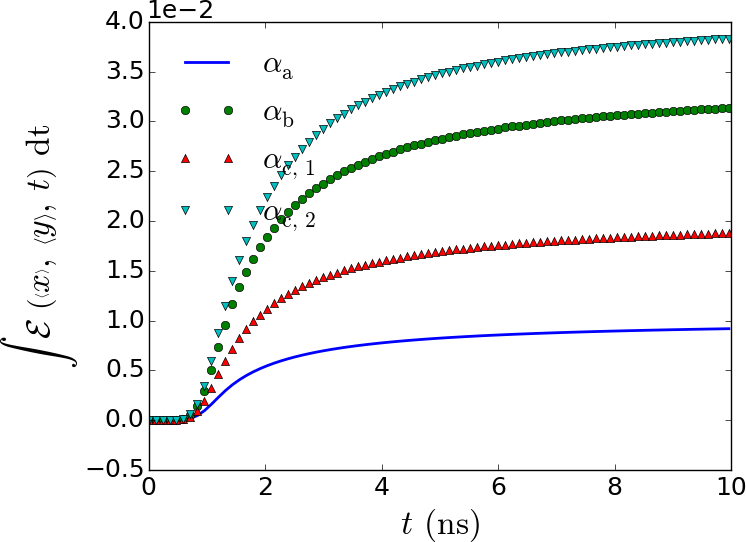}\protect\caption{
 \label{fig:Fig-8} The cumulative energy density as a function of time in the ABL region for the different profiles  at $\theta=63.4^{\circ}$.
 $\alpha_{\text{c},2}$ shows the largest energy transfer to the ABL.}
\end{figure}

\begin{figure}[tbh]
\centering{}\includegraphics[width=0.9\columnwidth]{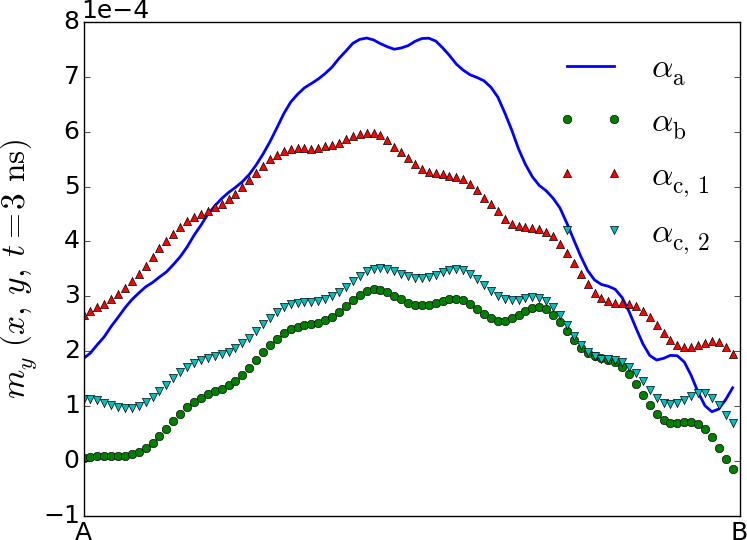}\protect\caption{
 \label{fig:Fig-9}The variation of the magnetization along the wavefront (red
line in  \autoref{fig:Fig-7}; simulations were run for $\alpha_b$ and $\alpha_{c,1}$ also). 
The $\alpha_{\text{b}}$ and $\alpha_{\text{c,2}}$ profiles show least reflections from the ABL.}
\end{figure}

  \autoref{fig:Fig-8} shows the cumulative energy density, from Eq. (\autoref{eq:5}), in the ABL for the different profiles at $\theta=63.4^{\circ}$.
$\alpha_{\text{c},2}$ leads to maximum absorption of SWs in the ABL and thus is the most efficient of all the profiles we have considered.
\autoref{fig:Fig-9} shows the magnetization scanned along a wavefront of the reflected wave, which is 
shown by the red line in  \autoref{fig:Fig-7}. Here too the amplitude of the reflected SW beam is low for $\alpha_{\text{c,2}}$.
\section{Summary and conclusions}

Reducing unwanted reflections from boundaries is
important for accurate simulations of magnonic devices. Shorter ABLs
with abrupt changes in $\alpha$ can cause spurious artifacts.
We calculated the return loss introduced in a permalloy stripe due to
the SWs normally incident on a ABL, using a transmission line model.
The parabolic damping profile yields a higher return loss, $1.5\,\text{dB}$ higher than an abrupt ABL.

We then considered SWs obliquely incident on the ABL at different angles of incidence. 
Even at a large incidence angle of $63.4^{\circ}$, the parabolic profile $\alpha_{\text{c},\,2}$ causes minimal reflections
and leads to the largest SW energy transfer to the ABL.
The performance of the tan hyperbolic damping profile is comparable to that of the parabolic profile.
Yet we urge the micromagnetic community to adopt the latter so as to align ourselves with the established
use of PMLs in FDTD simulations.

Example scripts to analyze the data, as
well as raw data for the figures, are available in the associated
electronic supplementary material \cite{Git_codes}.

\section*{Acknowledgments}
The authors would like to thank Malathi M., Manas Srivastava and Rajavardhan T 
for fruitful discussions and the High Performance Computing Centre
at IIT Madras for the use of their GPU cluster. This work was supported in part by the Department of 
Science and Technology, Government of India sanction number SB/S3/EECE/011/2014 (IITM).

\section*{Appendix A. Standing waves on a transmission line}
Consider a lossy transmission line extending from $x=0$ to $x=x_{0}$ .
If a wave is launched on a lossy transmission line towards the right at $x=0$ and the line is
terminated by an unmatched load, standing waves will be formed on
the line at steady state [Pozar 1997]. For small signal magnetization, assuming
linear systems, the standing waves are written as a sum of incident
and reflected waves as
\begin{eqnarray*}
  \begin{aligned}
    m_{\text{tot}}\left(x\right) & = & m^{+}e^{-\zeta x}e^{-j\beta x}+m^{-}e^{+\zeta x}e^{+j\beta x},
  \end{aligned}  
\end{eqnarray*}
where $m^{+}$, $m^{-}$, $\zeta$ and $\beta$ are the maximum amplitude
of incident and reflected waves, the loss per unit length and the
propagation constant of the wave respectively. We then have
\begin{eqnarray*}
  \begin{aligned}
    m_{\text{tot}}\left(x\right) & = & m^{+}\left[e^{-\zeta x}e^{-j\beta x}+\Gamma e^{+\zeta x}e^{+j\beta x}\right],
  \end{aligned}
\end{eqnarray*}
where $\Gamma=\left|\Gamma\right|e^{j\phi}$ is the reflection coefficient at the load end $x=x_{0}$.
The real part of $m_{\text{tot}}\left(x\right)$ is
\begin{eqnarray*}
  \begin{aligned}
    m\left(x\right) & = & m^{+}\left[e^{-\zeta x}\cos\beta x+\left|\Gamma\right|e^{+\zeta x}\cos\left(\beta x+\phi\right)\right].
  \end{aligned}  
\end{eqnarray*}

\section*{Appendix B. Implementation of the ABL in Mumax3}

To assist the interested reader, we reproduce the MuMax3 code for
setting the parabolic damping profile at the edge of the stripe in
\autoref{fig:Fig-1} (a).
We define each cell in the ABL
as a region and set the parabolic damping in it.
We define the start and stop damping values, and the range
of $x$ values.
\begin{verbatim}
alstart := 0.0	//alpha at start of ABL
alstop := 1.0	//alpha at stop of ABL
xstart := 3800	//x at start of ABL in nm
xstop := 4000	//x at stop of ABL in nm
n := 2		//Polynomial order
a := (alstop-alstart)/	//Polynomial coefficient
    (Pow((xstop-xstart), nxp))
cX := 5e-9 //Cellsize along x
NB := ((xstop-xstart)*1e-9)/cX	//No. of cells in ABL

//Set the damping cellwise
for i :=0; i<NB; i++{
xcurr := xstart*1e-9 + i*cX
DefRegion(i, xrange(xcurr, xcurr + cX))
alp := a*Pow((xcurr*1e9) - xstart, n)
alpha.setregion(i, alp)
} 
\end{verbatim}

\section*{References}
\biboptions{numbers,sort&compress}
\bibliographystyle{elsarticle-num}
\bibliography{paper}

\begin{thebibliography}{10}
\expandafter\ifx\csname url\endcsname\relax
  \def\url#1{\texttt{#1}}\fi
\expandafter\ifx\csname urlprefix\endcsname\relax\def\urlprefix{}\fi
\expandafter\ifx\csname href\endcsname\relax
  \def\href#1#2{#2} \def\path#1{#1}\fi

\bibitem{Bance/Schrefl/Hrkac/2008}
S.~Bance, T.~Schrefl, G.~Hrkac, A.~Goncharov, D.~A. Allwood, J.~Dean,
  Micromagnetic calculation of spin wave propagation for magnetologic devices,
  J. Appl. Phys. 103~(7) (2008) 07E735.
\newblock \href {http://dx.doi.org/10.1063/1.2836791}
  {\path{http://dx.doi.org/10.1063/1.2836791}}.

\bibitem{Peng201757}
Y.~Peng, G.~Zhao, F.~Morvan, S.~Wu, M.~Yue, Dynamic micromagnetic simulation of
  the magnetic spectrum of permalloy nanodot array with vortex state, J. Magn.
  Magn. Mater 422 (2017) 57--60.
\newblock \href {http://dx.doi.org/10.1016/j.jmmm.2016.08.060}
  {\path{http://dx.doi.org/10.1016/j.jmmm.2016.08.060}}.

\bibitem{Magnonics_Review}
V.~V. Kruglyak, et~al., Magnonics, J. Phys. D: App. Phys. 43 (2010) 264001.
\newblock \href {http://dx.doi.org/10.1088/0022-3727/43/26/264001}
  {\path{http://dx.doi.org/10.1088/0022-3727/43/26/264001}}.

\bibitem{SKKim/2010}
S.~K. Kim, Micromagnetic computer simulations of spin waves in nanometre-scale
  patterned magnetic elements, J. Phys. D: App. Phys. 43~(26) (2010) 264004.
\newblock \href {http://dx.doi.org/10.1088/0022-3727/43/26/264004}
  {\path{http://dx.doi.org/10.1088/0022-3727/43/26/264004}}.

\bibitem{Li201649}
Z.~Li, M.~Wang, Y.~Nie, D.~Wang, Q.~Xia, W.~Tang, Z.~Zeng, G.~Guo, Spin-wave
  propagation spectrum in magnetization-modulated cylindrical nanowires, J.
  Magn. Magn. Mater 414 (2016) 49 -- 54.
\newblock \href {http://dx.doi.org/10.1016/j.jmmm.2016.04.057}
  {\path{http://dx.doi.org/10.1016/j.jmmm.2016.04.057}}.

\bibitem{silvani2017spin}
R.~Silvani, M.~Kostylev, A.~Adeyeye, G.~Gubbiotti, Spin wave filtering and
  guiding in {P}ermalloy/iron nanowires, J. Magn. Magn. Mater (2017) --\href
  {http://dx.doi.org/10.1016/j.jmmm.2017.03.046}
  {\path{http://dx.doi.org/10.1016/j.jmmm.2017.03.046}}.

\bibitem{standard_micromagnetic_problem_Sw_dispersion}
G.~Venkat, D.~Kumar, M.~Franchin, O.~Dmytriiev, M.~Mruczkiewicz, H.~Fangohr,
  A.~Barman, M.~Krawczyk, A.~Prabhakar, Proposal for a standard micromagnetic
  problem: Spin wave dispersion in a magnonic waveguide, IEEE Trans. Magn.
  49~(1) (2013) 524--529.
\newblock \href {http://dx.doi.org/10.1109/TMAG.2012.2206820}
  {\path{http://dx.doi.org/10.1109/TMAG.2012.2206820}}.

\bibitem{Role_bound_2014}
M.~F{\"a}hnle, A.~Slavin, R.~Hertel, Role of the sample boundaries in the
  problem of dissipative magnetization dynamics, J. Magn. Magn. Mater 360
  (2014) 126--130.
\newblock \href {http://dx.doi.org/10.1016/j.jmmm.2014.02.031}
  {\path{http://dx.doi.org/10.1016/j.jmmm.2014.02.031}}.

\bibitem{Role_of_boundary_2013}
O.~Dmytriiev, V.~V. Kruglyak, M.~Franchin, H.~Fangohr, L.~Giovannini,
  F.~Montoncello, Role of boundaries in micromagnetic calculations of magnonic
  spectra of arrays of magnetic nanoelements, Phys. Rev. B 87 (2013) 174422.
\newblock \href {http://dx.doi.org/10.1103/PhysRevB.87.174422}
  {\path{http://dx.doi.org/10.1103/PhysRevB.87.174422}}.

\bibitem{berkov:08Q701}
D.~V. Berkov, N.~L. Gorn, Micromagnetic simulations of the magnetization
  precession induced by a spin-polarized current in a point-contact geometry
  (invited), J. Appl. Phys. 99~(8) (2006) 08Q701.
\newblock \href {http://dx.doi.org/10.1063/1.2151800}
  {\path{http://dx.doi.org/10.1063/1.2151800}}.

\bibitem{Consolo/abrupt_damping/2007}
G.~Consolo, L.~Lopez-Diaz, L.~Torres, B.~Azzerboni, Boundary conditions for
  spin-wave absorption based on different site-dependent damping functions,
  IEEE Trans. Magn. 43~(6) (2007) 2974--2976.
\newblock \href {http://dx.doi.org/10.1109/TMAG.2007.893124}
  {\path{http://dx.doi.org/10.1109/TMAG.2007.893124}}.

\bibitem{Dvornik11a}
M.~Dvornik, A.~N. Kuchko, V.~V. Kruglyak, Micromagnetic method of s-parameter
  characterization of magnonic devices, Journal of Applied Physics 109~(7)
  (2011) 07D350.
\newblock \href {http://dx.doi.org/DOI:10.1063/1.3562519}
  {\path{http://dx.doi.org/DOI:10.1063/1.3562519}}.

\bibitem{SW-abrupt-damping-1}
B.~V. de~Wiele, F.~Montoncello, A continuous excitation approach to determine
  time-dependent dispersion diagrams in 2{D} magnonic crystals, J. Phys. D:
  App. Phys. 47~(31) (2014) 315002.
\newblock \href {http://dx.doi.org/10.1088/0022-3727/47/31/315002}
  {\path{http://dx.doi.org/10.1088/0022-3727/47/31/315002}}.

\bibitem{Zhang_2015}
X.~Zhang, M.~Ezawa, D.~Xiao, G.~P. Zhao, Y.~Liu, Y.~Zhou, All-magnetic control
  of skyrmions in nanowires by a spin wave, Nanotechnology 26~(22) (2015)
  225701.
\newblock \href {http://dx.doi.org/10.1109/int-mag.2015.7157728}
  {\path{http://dx.doi.org/10.1109/int-mag.2015.7157728}}.

\bibitem{Xia_2016}
J.~Xia, X.~Zhang, M.~Yan, W.~Zhao, Y.~Zhou, Spin-{C}herenkov effect in a
  magnetic nanostrip with interfacial {D}zyaloshinskii-{M}oriya interaction,
  Scientific Reports 6~(25189).
\newblock \href {http://dx.doi.org/10.1038/srep25189}
  {\path{http://dx.doi.org/10.1038/srep25189}}.

\bibitem{PML_criteria}
X.~L. Travassos, S.~L. Avila, D.~Prescott, A.~Nicolas, L.~Krahenbuhl, Optimal
  configurations for perfectly matched layers in {FDTD} simulations, IEEE
  Trans. Magn. 42~(4) (2006) 563--566.
\newblock \href {http://dx.doi.org/10.1109/TMAG.2006.871471}
  {\path{http://dx.doi.org/10.1109/TMAG.2006.871471}}.

\bibitem{MuMax3}
A.~Vansteenkiste, J.~Leliaert, M.~Dvornik, M.~Helsen, F.~Garcia-Sanchez,
  B.~Van~Waeyenberge, The design and verification of {M}u{M}ax3, AIP Advances
  4~(10) (2014) 107133.
\newblock \href {http://dx.doi.org/10.1063/1.4899186}
  {\path{http://dx.doi.org/10.1063/1.4899186}}.

\bibitem{Git_codes}
G.~Venkat, H.~Fangohr, A.~Prabhakar, Post processing codes for paper on
  absorbing boundary layers for spin wave micromagnetics, {G}itHub.
\newblock \href {http://dx.doi.org/10.5281/zenodo.161326}
  {\path{http://dx.doi.org/10.5281/zenodo.161326}}.

\bibitem{Landau35}
L.~Landau, L.~Lifshitz, Theory of the dispersion of magnetic permeability in
  ferromagnetic bodies, Phys. Zeit. Sowjetunion 8 (1935) 153.
\newblock \href {http://dx.doi.org/10.1016/b978-0-08-010586-4.50023-7}
  {\path{http://dx.doi.org/10.1016/b978-0-08-010586-4.50023-7}}.

\bibitem{Franchin09}
M.~Franchin, \href{http://eprints.soton.ac.uk/id/eprint/161207}{Multiphysics
  simulations of magnetic nanostructures}, Ph.D. thesis, Univ. of Southampton
  (2009).
\urlprefix\url{http://eprints.soton.ac.uk/id/eprint/161207}

\bibitem{Kalinikos/1980}
B.~Kalinikos, Excitation of propagating spin waves in ferromagnetic films, IEE
  Proc. 127~(1) (1980) 4.
\newblock \href {http://dx.doi.org/10.1049/ip-h-1.1980.0002}
  {\path{http://dx.doi.org/10.1049/ip-h-1.1980.0002}}.

\bibitem{pozar1997microwave}
D.~Pozar, Microwave Engineering, 2nd Edition, Wiley, 1997.

\bibitem{Goos-Hanchen-SWs}
P.~Gruszecki, Y.~S. Dadoenkova, N.~N. Dadoenkova, I.~L. Lyubchanskii,
  J.~Romero-Vivas, K.~Y. Guslienko, M.~Krawczyk, Influence of magnetic surface
  anisotropy on spin wave reflection from the edge of ferromagnetic film, Phys.
  Rev. B 92 (2015) 054427.
\newblock \href {http://dx.doi.org/10.1103/physrevb.92.054427}
  {\path{http://dx.doi.org/10.1103/physrevb.92.054427}}.

\end{thebibliography}

\end{document}